# Magnetoresistive sensors based on the elasticity of domain walls

Xueying ZHANG[1,2,3], Nicolas VERNIER[3], Zhiqiang CAO[1,2], Qunwen LENG[1], Anni CAO[1,2], Dafiné RAVELOSONA[3] and Weisheng ZHAO[1,2]*

[1] Beihang-Goertek Joint Microelectronics Institute, Qingdao Research Institute, Beihang University, Qingdao, China

[2] Fert Beijing Institute, BDBC, School of Electronic and Information Engineering, Beihang University, Beijing, China

[3] Centre for Nanoscience and Nanotechnology, University Paris-Saclay, 91405 Orsay, France

*Weisheng.zhao@buaa.edu.cn

## Abstract

Magnetic sensors based on the magnetoresistance effects have a promising application prospect due to their excellent sensitivity and advantages in terms of the integration. However, competition between higher sensitivity and larger measuring range remains a problem. Here, we propose a novel mechanism for the design of magnetoresistive sensors: probing the perpendicular field by detecting the expansion of the elastic magnetic Domain Wall (DW) in the free layer of a spin valve or a magnetic tunnel junction. Performances of devices based on this mechanism, such as the sensitivity and the measuring range can be tuned by manipulating the geometry of the device, without changing the intrinsic properties of the material, thus promising a higher integration level and a better performance. The mechanism is theoretically explained based on the experimental results. Two examples are proposed and their functionality and performances are verified via micromagnetic simulation.

## I. Introduction

Magnetic sensors are of great importance for the intelligent world, especially in the field of the positioning, the navigation and the automatic control etc. A variety of effects related to the magnetism were used to fabricate the magnetic sensors[1–3], such as Hall sensors, Anisotropic Magnetoresistance (AMR) sensors[4,5], fluxgate sensors[6] etc. Magnetic sensors based on the Giant Magnetoresistance (GMR) or Tunneling Magnetoresistance (TMR) effects have been intensively studied in the past thirty years[7–14]. Thanks to the high sensitivity[15,16], scaled size[17,18] and compatibility with standard CMOS technologies[19], GMR sensors and TMR sensors have obtained great success in many applications[10], for example, in the detection of ultra-weak fields[20], in the position, speed and angle detection in automotive applications[21,22], in electrical current sensing[23] and in biomedicine[24–28] etc. When a probe equipped with a spin valve or a magnetic tunnel junction (MTJ) is placed in an external field, the magnetization of the free layer will tilt away from the easy axis. The tilting angle is determined by the competition between the effective anisotropy energy and the Zeeman energy due to the external field. The anisotropy can be achieved by the demagnetizing anisotropy[29], by the exchange bias[1,30] or straightforwardly, by a permanent magnet[31]. Compared with GMR sensors, TMR sensors promise a higher detectivity, a higher stability and lower power consumption. However, the high coercivity and low linearity for a large range remain two competing serious problems for the further improvement of the performance of TMR sensors[2]. For the in-plane anisotropy assured by the demagnetizing field, the strength of anisotropy can be tuned via manipulating the geometry of the device[29]. Therefore, sensors with different geometry, which are capable to detect the in-plane field with various range could be fabricated on the same chip. However, this regulatory method does not work for the device to detect the perpendicular field or for devices with perpendicular magnetic anisotropy (PMA) in the free layer. In some literatures[18], it was proposed to tune performances such as the sensitivity and the sensing range of sensors via modulating the PMA in the free layer, which can be controlled through the thickness of the free layer. However, this method cannot be used to fabricate devices with different performance in the same chip, i.e, not beneficial for the integration.

Therefore, to improve the integration level and the performance of the 3-axis sensor, integrated sensors capable to detect the magnetic field with variable sensitivity and measuring range is being desired.

In this paper, we propose a novel mechanism based on the elasticity of DWs for the design of magnetoresistive sensors: rather than probing the magnetic field via detecting the tilting angle of the complete free layer, we probe the field induced DW expansion in the free layer of the spin valve or in the MTJ. Performances, such as the measuring range and sensitivity of this type of sensors can be tuned via manipulating the geometry and the size of devices. Moreover, since the propagation field of DWs is much lower than the effective anisotropy field, problems caused by the high coercivity could be overcome. Two examples based on this mechanism are given and are verified via micromagnetic simulation. Devices based on this mechanism may greatly improve the performance and integration level of the 3-axis magnetic sensor.

## II. Concept and mechanism

In some soft magnetic materials, such as the CoFeB-MgO thin films, the intrinsic pinning field has been found to be ultra-weak[32]. In this case, the elasticity (or the surface tension) of the DW becomes the dominant factor for the DW behavior. Magnetic DW can be seen as an elastic membrane with surface energy γ (ref 33). The stable state of the DW is mainly determined by the competition between the elasticity and the Zeeman force due to the external field. For example, in our previous studies, we found that a semi-circular domain bubble could spontaneous collapse under the Laplace pressure due to the surface tension[34], as shown in Fig. 1. A small external field was required to stabilize the domain bubble. Moreover, the stabilizing field $H_{st}$ is approximately linear to the inverse of the radius R of the semi bubble. This is explained by the balance between the Laplace pressure, the Zeeman pressure due the external field and the demagnetizing field $H_d$, as shown in Fig. 1(d):

$$H_{st} = \frac{\gamma}{\mu_0 M_S R} - H_d \qquad (1)$$

where μ0 is the permeability of vacuum, Ms is the saturation magnetization.

Another example where the DW surface tension plays a dominant role is the pinning and depinning effect in the artificial geometry. We found that when a DW passes a Hall cross or when it is injected into a larger area from a narrow wire, it would be pinned at the intersection. The threshold field to depin the DW is also linearly dependent on the inverse of the wire width. This is also explained by the balance between the elasticity of the DW, which hinders the DW expansion, and the driving force from the external field.

Note that the propagation of the DW results in a local change of the magnetization. In the free layer of an MTJ, this change can be accurately detected through the change of the magnetoresistance. Interestingly, the structure near the free layer of the generally used MTJ, composed of MgO/CoFeB/heavy metal, exerts many excellent properties, which facilitate the accurate DW manipulation, for example, the high PMA[35] and the ultralow intrinsic pinning fields[32]. We give in following an example to show the principle to measure the magnetic field using the elasticity of DWs and the method to tune the performance of sensors through manipulating the geometry and the size of devices.

### III. Device design and simulations

The structure of one example we proposed based on the above mechanism is shown in Fig.2. The free layer of two or more MTJs are connected via wire bridges while he pinned layer of MTJs are isolated. Free layers have a PMA and magnetizations of pinned layers are perpendicularly initialized on the same direction.

Before working, opposite current pulses should be applied so that the magnetization of the free layer of two adjacent MTJs are initialized to opposite directions via the spin transfer torque. A DW will be created in the bridge. When the device is put in an external field, the DW will be moved in either of the two directions, depending on the direction of the external field. When the DW arrives at the connection between the bridge and one of the free layer (i.e. the entrance), it will be pinned, since the further expansion means a larger DW surface and thus a raised DW surface energy. As described by Eq. (1), in equilibrium, the DW will be of a circular arc shape and the radius can be given as follows,

$$R = \frac{\gamma}{\mu_0 M_S (H_d + H_{ext})} \tag{2}$$

Here, the demagnetizing field $H_d$ is determined by the magnetic state and the structure of the device, varying with $H_{ext}$. Other parameters are all intrinsic. Therefore, the radius of the DW arc is solely dependent on $H_{ext}$.

For a given width w of the bridge, the area reversed in the free layer caused by the expansion of the DW can be expressed as,

$$S = R^2 \sin^{-1}\left(\frac{w}{2R}\right) - \frac{w}{4}\sqrt{4R^2 - w^2} \tag{3}$$

The change of this area leads to the change of the magnetoresistance. In this way, the external field is quantified through the resistance of the MTJ. Since the readout of the magnetoresistance can be realized with very low current density, the effect of this current on the behavior of the DW can be neglected.

In order to verify the functionality of this proposed device, the response of the magnetization in the free layer of the device versus external fields was simulated via micromagnetic using Mumax[36]. As shown in Fig.3, a DW was set in the middle of the bridge as the initial state. Then a magnetic field was applied. After the DW was pinned at the entrance, the magnitude of the magnetic field is increased gradually and the corresponding DW states were shown in Fig.3. We can see that the radius of the DW decreases and that the DW expands into the MTJ as the field increases. After the external field is removed, DW can come back to the bridge owing to its elasticity.

During the simulation, the perpendicular component of the average magnetization $\bar{M}_z$ of the free layer was extracted and was plotted as a function of the external field, as shown in Fig. 4(a). Since the resistance of the spin valve or the MTJ is directly determined by $\bar{M}_z$, these figures can be seen as an indicator of the resistive response of the devices to the measured field. From Fig4(a), we can see that the response of the magnetoresistance of the two MTJs to the external field are complementary, because the initial magnetization are set to be opposite. Therefore, a couple of MTJs must work synergistically to realize the detection of the magnetic field in opposite directions.

From the comparison of the figure in left and right in Fig. 4(a), we can see that the device with larger size (here, the size is characterized by the width of the bridge) has a better sensitivity but the measuring range is relatively small; vice versa. In Fig.4 (b), the measuring range and the sensitivity of the device versus the size of devices were plotted. Here, the sensitivity is defined as the rate of the change of the average perpendicular magnetization (normalized from -1 to 1) with respect to the change of the external field.

In order to verify the response of the device to the alternating field, we simulated the change of the magnetization in a 500nm wide device when a sinusoidal field with an amplitude of 12.5 mT and a frequency of 1M Hz was applied. The perpendicular component of the average magnetization is plotted in figure 4 (c).

## IV. Discussions

When the DW is blown from the bridge into the free layer, the radius of the DW arc decreases as $H_{ext}$ increases. When $H_{ext}$ reaches a threshold value, the diameter of the arc decreased to the value of the width of the bridge and DW will be depinned, leading the entire reversal of the free layer. Therefore, the depinning field is the upper limit of the range of the device[34],

$$H_{max} = \frac{2\gamma}{\mu_0 M_S w} - H_d \tag{3}$$

It can be seen that the measuring range of the device is mainly determined by the width of the bridge. For a narrower bridge, the measuring range will be raised. According to our previous experiments, in a Ta/CoFeB/MgO structure, the depinning field when injecting a DW from a 200 nm wire into a larger area is about 20mT (ref 34). If the size of the device scales down to tens of nanometers, a measuring range of hundreds mT is expected. While for a larger bridge, the measuring range decrease but the sensitivity increases, as shown in Fig.4 (b).

Simulation results have shown that these devices have a satisfactory performance for the detection of the alternating field with a frequency of MHz. In fact, the response speed is limited by the DW motion velocity in the free layer. According to our measurements (see supplementary information) in the Ta/CoFeB/MgO film, the DW motion velocity can reach more than 5m/s soon when the applied field $H_{ext}$ exceeds the intrinsic pinning field. Supposing that in a device of hundreds nm, DW can reach the ready status in less than 100 ns. For a lower external field, DW motion velocity is dominated by the intrinsic defects and decrease rapidly, obeying the creep law. Still, a movement of 100nm can be achieved in less than 1ms when the measured field decreases to 0.5 mT, according to our experimental results (See supplementary information). In addition, the length of the bridge connecting the free layer of two-coupled MTJs is also related to the response speed of the device when the measuring field changes the sign. By further decreasing the intrinsic pinning field of the free layer or reducing the length of the bridge, the sensitivity and the response speed of the device could be further improved.

Various devices with different size and geometry, thus with different performance (e.g. measuring range, sensitivity etc.) can be fabricated in the same chip based on this mechanism. No modulation of the intrinsic properties of the film, such as the PMA, is required. Therefore, magnetic sensors based on the elasticity of DWs promise a higher integration level and a better performance.

Note that the above device is only an example to demonstrate the concept we proposed and to verify the performance of this type of sensors. The monotonous and reversible response of the elastic DW to the magnetic field provides a new aspect to design sensors. Another example based on the asymmetric DW motion in the magnetic wire with a gradient width, which is also associated with the elasticity of DWs, is given in the supplementary information.

## V. Conclusions

In this work, we proposed a novel mechanism for the measurement of the magnetic field based on the elasticity of DWs. An external magnetic field can cause the expansion of a circular DW in the free layer of a spin valve or an MTJ. Thanks to the elasticity of DWs, this expansion is monotonously dependent on the magnitude of external fields and is reversible. Since this expansion leads to the change of the magnetization in the free layer, the external field can be quantified via the resistance of the device. Two examples based on this mechanism were proposed and verified with micromagnetic simulations. The device proposed shows a good performance on the measurement of the direct or alternating field. Performances such as the measuring range and the sensitivity can be tuned by manipulating the geometry and the size or device. The mechanism proposed here may greatly improve the integration level and performances of magnetic sensors.

## Acknowledgements

The authors would like to thank the supports by the projects from National Natural Science Foundation of China (No. 61571023 and 61627813), and the International Collaboration Project (No. 2015DFE12880 and No. B16001).

**Figures and Captions**

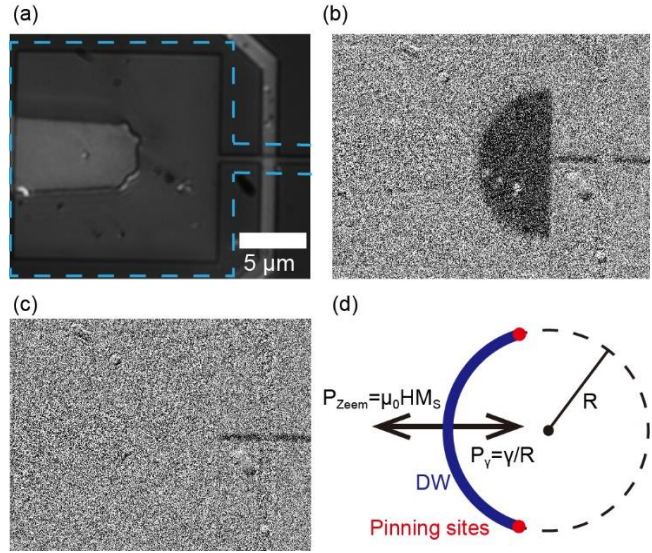

FIG.1 (a) Optical image of the magnetic structure tested (surrounded in blue): a magnetic square connected with a narrow wire, with Ta/CoFeB/MgO multi-layers structure; (b) A semicircular magnetic bubble with radius of 7 μm was created when an external field was applied. This bubble could be stabilized by an external field; (c) The domain bubble spontaneously collapsed when the external field was removed; (d) A circular DW arc pinned at the two terminals: Equilibrium can be achieved under the competition of Zeeman pressure and the Laplace pressure due to the DW surface tension (elasticity).

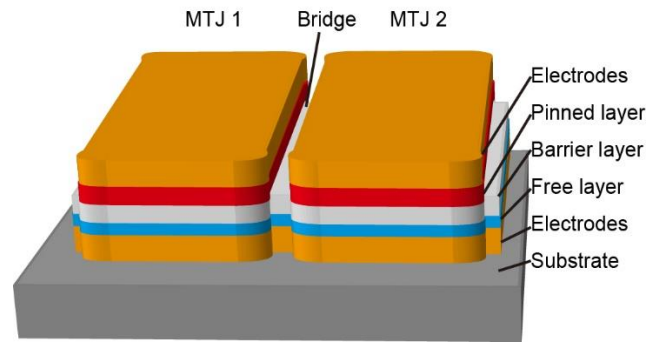

FIG.2 Structure of a sensor device proposed based on the elasticity of the DW. Two MTJs are connected via a bridge while the pinned layer and the upper electrodes are isolated.

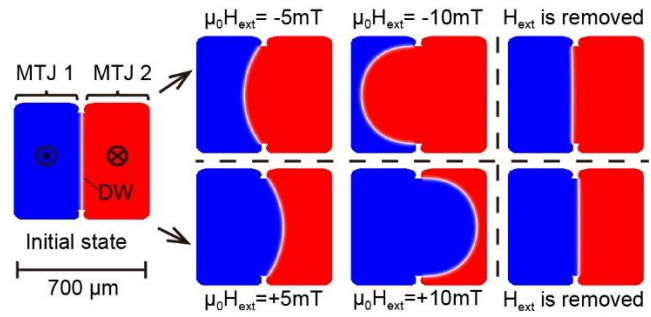

FIG.3 DW states in the free layer of two coupled MTJs: First, DW was set in the middle of the bridge of the device. Then an external perpendicular field of -5 mT (+5 mT), -10 mT (+10 mT) was applied and the stable state of the DW was snapshotted, respectively. At last, the external field was removed and the DW returned to the bridge due to its elasticity.

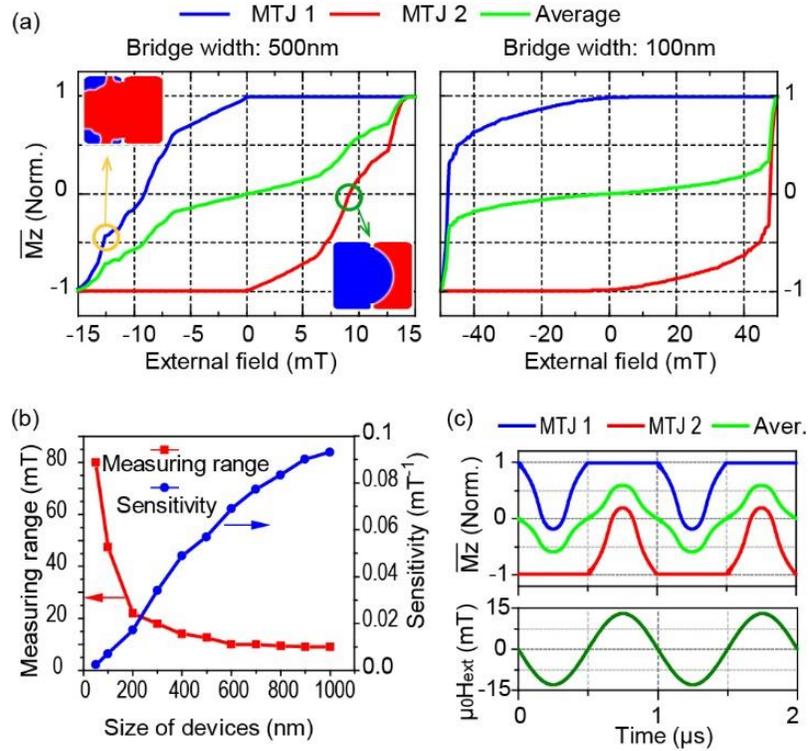

FIG.4 (a) The average magnetization in the free layer of the device as a function of the external field. The width of the bridge in the device simulated is 500 nm in the left and 100 nm in the right. Insets show the DW state in the free layer. When external field exceeds the measuring range (13mT, marked in yellow), domain bubble breaks. (b) The measuring range and the sensibility of devices as a function of the size of devices. (c) Average perpendicular magnetization in the free layer of MTJs when an alternating field is applied.